\documentclass[prb,twocolumn,superscriptaddress,floatfix,showpacs,amsmath,amssymb]{revtex4-2}

\usepackage{amssymb}
\usepackage{amsthm}
\usepackage{amsmath}
\usepackage[latin9]{inputenc}
\usepackage{graphicx,color}
\usepackage{blindtext}
\usepackage{booktabs}
\usepackage{caption}
\usepackage{textcomp}
\usepackage{epstopdf}
\usepackage{subfigure}
 \usepackage{multirow}
\usepackage{bm}
\usepackage{natbib}

\newcommand{\lno}{La$_4$Ni$_3$O$_{10}$}
\newcommand{\rno}{{\it R}$_4$Ni$_3$O$_{10}$}

\newcommand{\tmmt}{$T_{\rm MM}$}
\newcommand{\tmsone}{$T^{Bmab}_{\rm MM}$}
\newcommand{\tmstwo}{$T^{P2_1/a}_{\rm MM}$}
\newcommand{\chic}{$\chi_{\rm c}$}
\newcommand{\chiab}{$\chi_{\rm ab}$}

\newcommand{\mbfu}{\(\mu _{\rm B}/{\rm f.u.}\)}

\newcommand{\jmk}{J/(mol\,K)}
\newcommand{\etal}{\textit{et~al.}}

\newcommand{\Cp}{$c_{\rm p}$}

\begin{document}
	
\title{High-pressure crystal growth and investigation of the metal-to-metal transition of Ruddlesden-Popper trilayer nickelates La$_4$Ni$_3$O$_{10}$}

	\author{Ning Yuan}
	\affiliation{Kirchhoff Institute for Physics, Heidelberg University, 69120 Heidelberg, Germany}
	
	\author{Ahmed Elghandour}
	\affiliation{Kirchhoff Institute for Physics, Heidelberg University, 69120 Heidelberg, Germany}
		
	\author{Jan Arneth}
	\affiliation{Kirchhoff Institute for Physics, Heidelberg University, 69120 Heidelberg, Germany}

    \author{Kaustav Dey}
	\affiliation{Kirchhoff Institute for Physics, Heidelberg University, 69120 Heidelberg, Germany}
	
    \author{R\"udiger~Klingeler}
    \affiliation{Kirchhoff Institute for Physics, Heidelberg University, 69120 Heidelberg, Germany}

\begin{abstract}
Single crystals of Ruddlesden-Popper nickelates La$_4$Ni$_3$O$_{10}$ were grown by means of the floating-zone technique at oxygen pressure of 20~bar. Our results reveal the effects of the annealing process under pressure on the crystal structure. We present the requirements for crystal growth and show how a reported ferromagnetic impurity phase can be avoided. The different growth and post-annealing processes result in two distinct phases $P2_1/a$ and {\it Bmab} in which the metal-to-metal transitions occur at 152~K and 136~K, respectively.

\end{abstract}

\maketitle

\section{Introduction}

The recent discoveries of superconductivity in infinite-layer nickelates~\cite{Nature2019nickelate,PRL2020Nd1-x,PRL2020phasediagram,NC2020Nd1-x,NANOL2020pr,PRM2020pr,AM2021nickelate, ScienceAd2022LCNO} have further demonstrated the position of nickelates as model systems to discover and decipher novel aspects of correlated electron physics. It is also the electronic similarity of the Ni$^{1+}$ and Cu$^{2+}$ electronic configurations which renders nickelates prime analogues to the high-$T_{\rm C}$ superconducting cuprates and has raised a surge of interest to understand and modify the critical electronic features that determine electronic correlation and in particular superconductivity in nickelates. This has brought the Ruddlesden-Popper phases La$_{n+1}$Ni$_n$O$_{3n+1}$ into the focus in which the valence of the Ni-ions and the electronic ground state can be tuned (see, e.g., \cite{greenblatt1997ruddlesden,LaBollita2021,Jung2022,Pan2022,NP2017large}). Trilayer nickelates as reported at hand exhibit an unusual metal-to-metal transition (MMT)~\cite{JMMM2020LNO,PRB2020NNO,PRB2020PNO,PRB2020RNO-RK,PRM2020LNO,PRR2020RNO,PRB2001LNO} with intertwined charge and spin orders developing at \tmmt~\cite{NC2020intertwined}.

The crystal structure of Ruddlesden-Popper nickelates \rno\ is commonly described as an alternating arrangement of perovskite-like layers and rock-salt-like layers~\cite{greenblatt1997ruddlesden,JSSC1995LNO} and is often regarded as quasi-two-dimensional (2D). It exhibits a mixture of Ni$^{2+}$ and Ni$^{3+}$ ions with an average valence of +2.67~\cite{JMMM2020LNO}. The space group details at room temperature and ambient pressure are still controversial and the debated four different space groups are discussed specifically in Refs.~\cite{PRB2020RNO-RK,PRM2020LNO,PRB2018structure}. By means of high-resolution synchrotron and laboratory x-ray single-crystal diffraction studies Zhang~\etal\ concluded that the formation of the $P2_1/a$ (no.~14) and {\it Bmab} (no.~64) structures is closely related to the cooling rate after growth. The preparation of \lno\ compounds, especially of single crystals, is challenging due to the required synthesis atmosphere of 20 -- 30 bar oxygen pressure~\cite{greenblatt1997ruddlesden,PRM2020LNO} and the pronounced tendency of phase mixture which is reported to be closely associated with slight variations of the oxygen content~\cite{PRB2001LNO,JAP2000oxygen}.

In this work, single crystals of \lno\ were successfully grown by the high-pressure optical floating-zone method under 20~bar oxygen pressure. We investigate the different growth and post-annealing processes resulting in two different phases, i.e. $P2_1/a$ and {\it Bmab}, respectively, and report magnetic susceptibility and specific heat data. Our study shows sharp anomalies and marked anisotropy associated with the reported MMT in \lno. Using the here reported growth conditions avoids the formation of a previously reported ferromagnetic impurity phase.

\section{Materials and experimental methods}

\begin{figure}[b]
\centering
\includegraphics[width=1\columnwidth,clip]{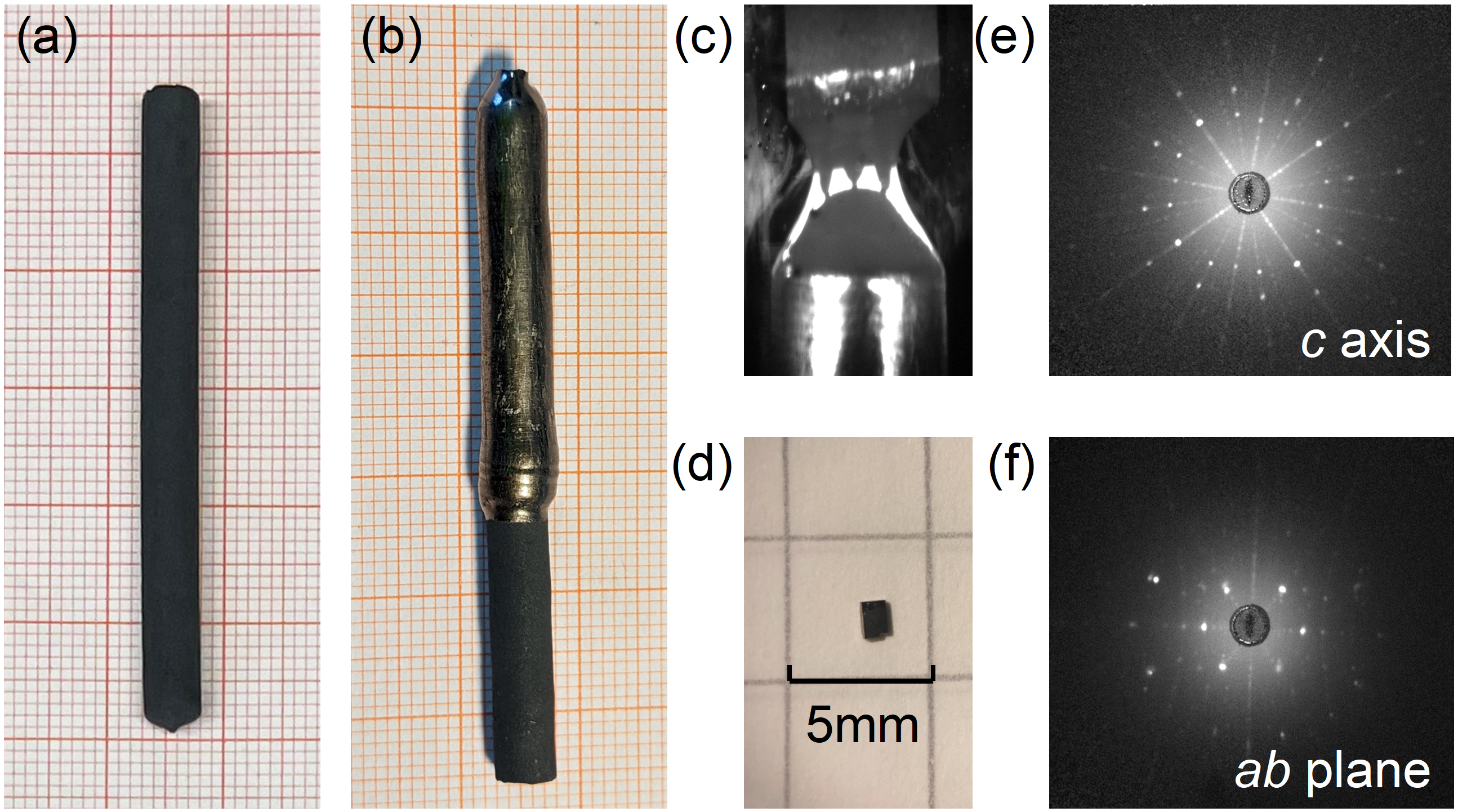}
\caption{Pictures of (a) \lno\ cylindrical polycrystalline rods, (b) the obtained \lno\ boule, (c) the melting-zone formed during the growth, and (d) the oriented and cut single crystal used for magnetic and calorimetric measurements. (e,f) Laue patterns of the \lno\ single crystal oriented along $c$ axis and $ab$ plane, respectively.} \label{Fig.1-Growth}
\end{figure}

Polycrystalline \lno\ was synthesized by a standard solid-state reaction. The raw materials La\(_2\)O\(_3\) (99.99$\%$, Sigma-Aldrich) and NiO (99.998$\%$, Alfa Aesar) powders were calcined at 900~$^\circ$C and 1000~$^\circ$C for 24~h to remove absorbed water. Stoichiometric amounts of the ingredients were mixed well in a mortar and calcined at 1050~$^\circ$C for 24~h (air flow, ambient) with several intermediate grindings. The powder obtained is reground, packed in a rubber tube and isotropically pressed at 60~MPa in order to produce cylindrical rods with a length of 5-6~cm and a diameter of 5~mm as shown in Fig.~\ref{Fig.1-Growth}a. The rods were annealed for 24~h at 1400~$^\circ$C. It is notable that according to the thermodynamic analysis of the La-Ni-O system~\cite{JAC2007stability,JAC2004thermodynamic}, only La\(_2\)NiO\(_4\) compounds can be synthesized at ambient pressure, and our experimental results are consistent with it. Single crystals of \lno\ were then successfully grown using the high-pressure optical floating-zone furnace (HKZ, SciDre)~\cite{neef2017high}. We employed a 5~kw Xenon arc lamp as the heat source, and 20~bar O\(_2\) atmosphere with an O\(_2\) flow rate maintained at 0.1~l/min. To improve the homogeneity of the melting zone, counter-rotation of the feed and seed rods at 10~rmp is necessary. The feed rod was pulled at 6~mm/h and the seed rod was pulled at 4~mm/h to maintain the zone stability. Using an in-situ temperature measurement by means of a two-color pyrometer~\cite{dey2019magnetic,hergett2019high}, the temperature of the melting zone during growth was determined to about 1650~$^\circ$C. After the initial growth we performed a rapid cooling of the melting zone to avoid the precipitation of oxygen. In addition to crystals obtained from pristine boules grown as described above and further on labelled `S1', we also applied a post-annealing procedure resulting in crystals labelled `S2'. The post-annealing was carried out in the HKZ furnace where the sample is held under the 20~bar oxygen pressure at 950~$^\circ$C for 2 hours, then quenched to room temperature following Ref.~\cite{PRM2020LNO}. The latter process was achieved by quickly removing the rod out of focus.

The phase purity and crystallinity of the resulting materials were studied by powder X-ray diffraction (XRD) and the back-reflection Laue method. XRD was performed at room temperature by means of a Bruker D8 Advance ECO diffractometer using Cu-K$\alpha$ radiation ($\lambda$ = 1.5418\,\AA). Data have been collected in the 2$\Theta$ range of 10 - 90$^\circ$ with 0.02$^\circ$ step-size. Laue diffraction was done on a high-resolution X-Ray Laue camera (Photonic Science). Magnetic studies in the temperature regime 1.8 - 350~K have been performed in a SQUID magnetometer (MPMS3, Quantum Design Inc.) following either field-cooled (FC) or zero-field-cooled (ZFC) protocols where the sample has been cooled down to lowest temperature in the actual measurement field or in zero magnetic field, respectively, before applying the external magnetic field at lowest temperature. Measurements of the specific heat have been performed in a Physical Properties Measurement System (PPMS, Quantum Design Inc.) utilizing a relaxation method.

\section{Results}

\subsection{Single crystal growth}

\begin{figure}[htb]
\centering
\includegraphics[width=\columnwidth,clip]{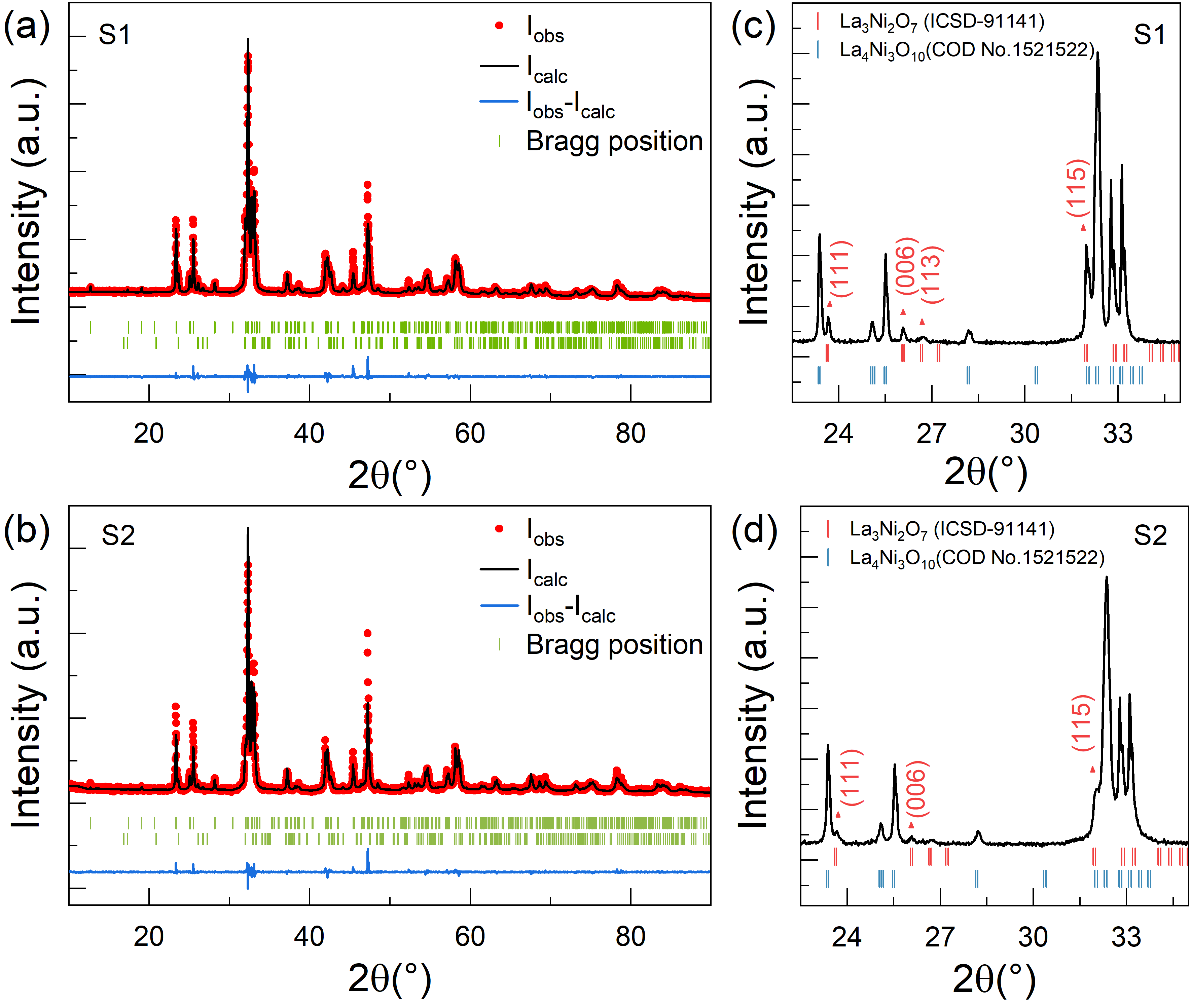}
\caption{Room temperature powder XRD patterns and corresponding Rietveld refinement~\cite{rodriguez2001introduction} of ground \lno\ single crystals grown at 20~bar O$_2$ pressure. S1 (a,c) and S2 (b,d) show results for pristine and annealed crystals, respectively (see the text). The observed diffraction pattern is shown in red, the calculated one in black, and the difference between them is shown in blue. The vertical green bars show the expected Bragg positions of \lno\ (COD no.~7237332~\cite{ling2000neutron}) and La\(_3\)Ni\(_2\)O\(_7\) (ICSD no.~91141~\cite{ling2000neutron}). The refinement converged to $R_{\rm p}$ = 15.9$\%$, $R_{\rm wp}$ = 16.2$\%$, $\chi^{2}$ = 5.7 for S1 and $R_{\rm p}$ = 27.2$\%$, $R_{\rm wp}$ = 24.8$\%$, $\chi^{2}$ = 3.7 for S2.}
\label{Fig.2}
\end{figure}

\lno\ single crystals were successfully grown under 20~bar oxygen pressure (see Fig.~\ref{Fig.1-Growth}b for a picture of the as-grown boule). In order to maintain a stable melting zone, the feed rod must be pulled faster than the seed rod; optimized feed and seed rod velocities chosen for the experiment were 6~mm/h and 4~mm/h, respectively. The melting zone formed during the successful growth is shown in Fig.~\ref{Fig.1-Growth}c. Using the same velocities for both rods resulted in the depletion of liquid in the melting zone, somehow similar as observed for LaNiO$_3$~\cite{dey2019magnetic}. The inconsistent velocity setting results in wider diameter of the as-grown single crystal compared to the feed (see Fig.~\ref{Fig.1-Growth}b) and in general entails a larger potential for cracking.

The procedure yields shiny boules from which single crystalline grains have been obtained, oriented, and cut (Fig.~\ref{Fig.1-Growth}b and d). Crystals were obtained from both the pristine boule (S1) and the annealed one (S2). Figures~\ref{Fig.1-Growth}d-f show the oriented and cut crystal S2 as well as back-reflection Laue images obtained along the $c$ axis and the $ab$ plane, respectively. Similar sample dimension and Laue patterns have been obtained for crystal S1.

Powder X-ray diffraction was performed on several ground single crystals taken from the very vicinity of the oriented single crystal bulks in order to study phase purity of the grown single crystals. The resulting room temperature XRD pattern as well as the Rietveld refinements to the data presented in Fig.~\ref{Fig.2}a,b indicate that the main phase is \lno . In addition, we observe a few Bragg peaks that do not correspond to \lno\ phases but are assigned to La\(_3\)Ni\(_2\)O\(_7\) impurities. We find that the annealing procedure under oxygen pressure leads to a notable reduction of the impurity phase, as shown in Fig.~\ref{Fig.2}c,d.

\subsection{Magnetization}

\begin{figure}[tb]
\centering
\includegraphics [width=0.9\columnwidth,clip] {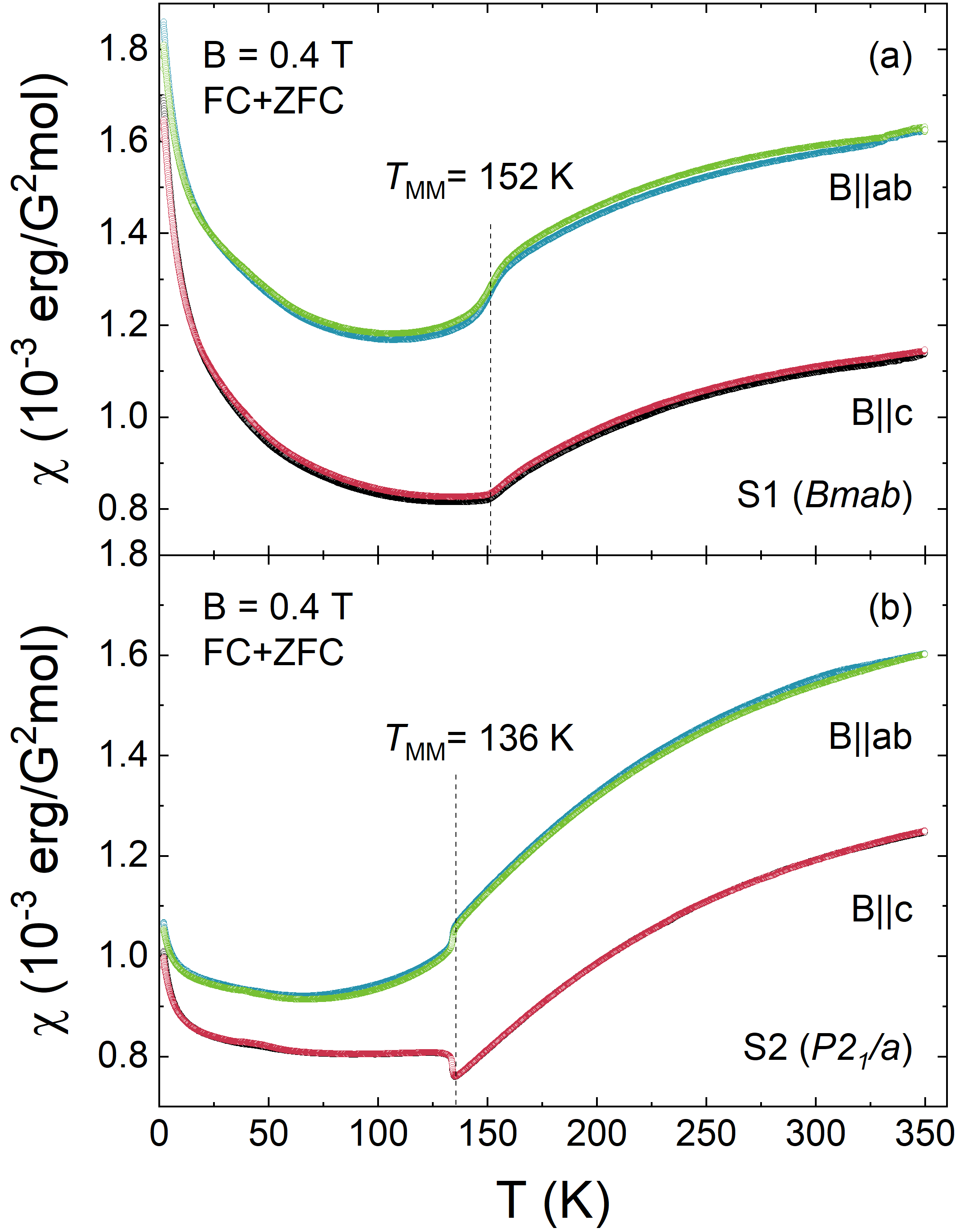}
\caption{Static magnetic susceptibility $\chi = M/B$ vs. $T$ measured at $B=0.4$~T of (a) the pristine crystal (S1:{\it Bmab}) and (b) the annealed crystal (S2:$P2_1/a$). For each field direction ($B||c$ and $B||ab$) both fc and zfc data are presented but do not show visible differences.}
\label{chiboth}
\end{figure}

The in-plane and out-of-plane magnetisation measurements performed on single crystal samples S1 and S2 shown in Fig.~\ref{chiboth} provide further information on the quality of the single crystals and on potential impurity phases. The main features are sharp jumps or kinks in the static susceptibility $\chi=M/B$ at \tmsone~=~152~K and \tmstwo~=~136~K for the pristine S1 and the annealed S2, respectively. This result is consistent with previous reports \cite{PRB2020NNO,PRB2020PNO,PRB2020RNO-RK,PRM2020LNO,PRR2020RNO,NC2020intertwined,NC2017electrondynamics} which imply that the MMT is characterised by an intertwined charge-magnetic ordering phenomenon.

We emphasize the absence of FC/ZFC hysteresis behaviour in the 1.8 to 350~K temperature interval (see Fig.~\ref{chiboth}); previously reported crystals show a pronounced anomaly in the susceptibility at around 50~K and strong differences between FC and ZFC susceptibilities, suggesting an unknown ferromagnetic component~\cite{PRM2020LNO}. This component is absent in our crystals which confirms its extrinsic nature~\footnote{Within the error bars of the experiment, we obtain an upper limit of the ferromagnetic component being 25 times less than in Ref.~\cite{PRM2020LNO}.}. The absence of ferromagnetic impurities in our crystals is further confirmed by isothermal magnetisation studies (see Fig.~\ref{MB} below).

Despite the challenges and controversy regarding the determination of crystal structure of \lno\ by powder XRD analysis~\cite{PRB2020RNO-RK,PRM2020LNO,PRB2018structure}, the magnetic susceptibility data not only imply phase purity with respect to magnetic impurity phases, but also strongly suggest that the pristine crystal S1 exhibits {\it Bmab} structure while S2 exhibits $P2_1/a$ structure. In particular, the very pronounced and sharp single anomalies exclude mixing of both phases in either of the single crystals. The very sharp jump in \chic\ at \tmmt\ in S2 in particular indicates excellent crystallinity of the $P2_1/a$ phase crystal. Zhang~\etal~\cite{PRM2020LNO} have identified the postgrowth cooling rate as a crucial parameter for obtaining a thermodynamically stable phase. The reported experiments on biphasic \lno\ crystals suggest that the {\it Bmab} structure can be transformed to $P2_1/a$. In our experiments we have performed rapid cooling of the as-grown boule as well as further high-pressure annealing. Our results show that the phase pure {\it Bmab} structure forms after rapid cooling while it completely transforms to $P2_1/a$ upon heat treatment confirming the latter being the thermodynamically stable phase.

\begin{table*}[h]
\centering
\caption{Fitting magnetisation data in Fig.~\ref{chiboth} and \ref{MB} (see the text). $M_s$ and $\chi$ are saturation magnetisation of the quasi-free spins and linear slope derived from $M(B,T=2~{\rm K})$. $C$, $\Theta$, and $\chi_0$ are the Curie constant, Weiss temperature and temperature-independent susceptibility from fitting $M(T)/B$ at $T<10$~K.}
\scalebox{0.8}{
\begin{tabular}{l|ccccc|ccccc}
  & $M_s^{\rm c}$ & \chic &$C$&$\Theta$&$\chi_0$& $M_s^{\rm ab}$ & \chiab &$C$&$\Theta$&$\chi_0$\\
  & \mbfu  & erg/(G$^2$mol) &erg\,K/(G$^2$mol)&K&erg/(G$^2$mol)& \mbfu & erg/(G$^2$mol) &erg\,K/(G$^2$mol)&K&erg/(G$^2$mol) \\
\hline
S1({\it Bmab})& $2.2\times 10^{-3}$ & $3.2\times 10^{-3}$ &$10\times 10^{-3}$&9&$8\times 10^{-4}$ & $1.8\times 10^{-3}$ & $3.8\times 10^{-3}$ &$6\times 10^{-3}$&7&$1.2\times 10^{-3}$\\
S2($P2_1/a$) & $7.8\times 10^{-4}$ & $2.2\times 10^{-3}$ &$1.1\times 10^{-3}$&3&$8\times 10^{-4}$& $8.2\times 10^{-4}$ & $7.7\times 10^{-4}$ &$7\times 10^{-4}$&3&$8\times 10^{-4}$\\
\end{tabular}\label{tab1}
}
\end{table*}

In general, the susceptibility data in Fig.~\ref{chiboth} show (1) the continuous decrease of $\chi$ upon cooling from 350~K towards \tmmt , with a minimum in $\chi_{\rm c}$ at \tmmt\ and a broad minimum in $\chi_{\rm ab}$ slightly below. (2) Pronounced anisotropy in the whole temperature range under study with \chiab /\chic = 1.42 (S1) respectively 1.28 (S2), at 350~K. (3) A Curie-Weiss-like (CW-like) increase of the magnetic susceptibility $\chi(T)$ at low temperatures which qualitatively corresponds to a Brillouin-like contribution to $M(B,T=2~{\rm K})$ (see Fig.~\ref{MB} below).

Notably, the seemingly CW-like behaviour of magnetic susceptibility in both the $P2_1/a$ (S1) and {\it Bmab} (S2) systems well below \tmmt\ (see Fig.~\ref{chiboth}) cannot be described well in terms of the extended CW law $\chi (T)=\frac{C}{T+\Theta}+\chi_0$, with $C$ the Curie constant, $\Theta$ the Weiss temperature, and a temperature independent term $\chi_0$, in an extended temperature regime. This implies that at least one or even all these parameters change upon cooling. In an attempt to quantify these parameters at low temperature by restricting the CW fit to $T<10$~K, the data are reasonably well described by the parameters listed in Table~\ref{tab1}. The small values of $\Theta$ imply the presence of very weakly coupled magnetic moments. When evaluating the Curie constant by assuming magnetic moments $S=1$ and $g\simeq 2$, the data suggest 0.4~\textperthousand/f.u. of such moments. The observed $\chi(T)$ however also implies that the number of localized moments, their magnetic interaction and/or $\chi_0$ change with temperature.

\begin{figure}[h]
\centering
\includegraphics [width=0.9\columnwidth,clip] {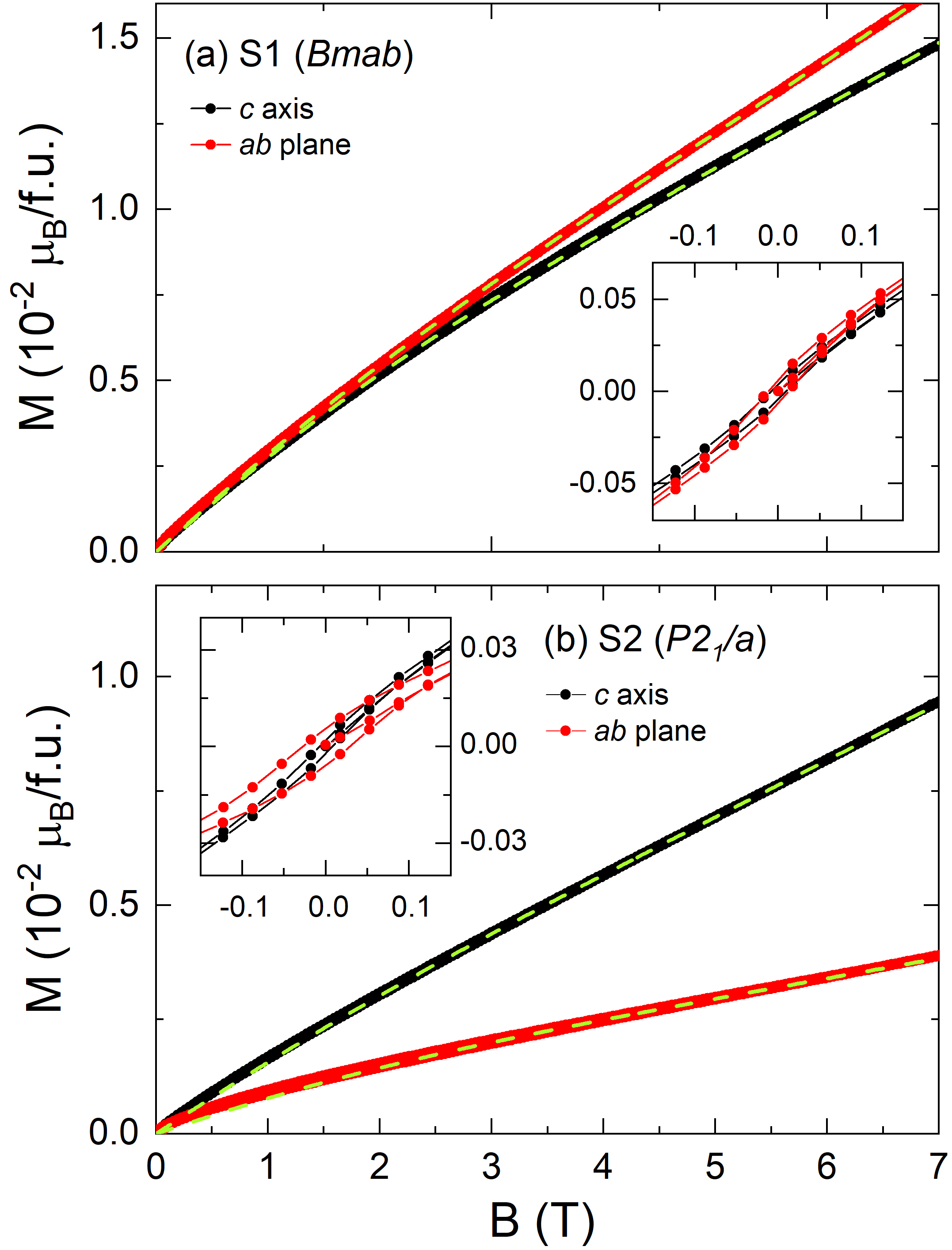}
\caption{Isothermal magnetization of (a) the pristine crystal (S1:{\it Bmab}) and (b) the annealed crystal (S2:$P2_1/a$), at $T=2$~K, for $B||c$ and $B||ab$ (both up- and down-sweeps are shown). Dashed lines extrapolate the linear high-field behaviour from which $\chi_{c}$ and $\chi_{ab}$ in Table~\ref{tab1} are derived. Insets: $M$ vs. $B$ for -0.15~T~$\leq B\leq$~0.15~T. For a full $M(B)$ curve ranging from -7~T to +7~T see the Supplemental Material~\cite{SM}.}
\label{MB}
\end{figure}

The isothermal magnetization (Fig.~\ref{MB}) shows rather linear behaviour in the high-field regime but also clear signatures of quasi-free spins as there is distinct right-bending in small fields. We attribute the linear behaviour to the response of the main magnetic phase. Considering these contributions, $M(B)$ at $T=2$~K may be described by
\begin{equation}
    M(B)=M_{\rm s}\times B_{S}(\frac{g\mu_\mathrm{B}SB}{k_BT})+\chi_0B.
\end{equation}
Here, $M_{\rm s}$ is the saturation magnetization of the quasi-free moments, $B_S$ the Brillouin function, $S$ spin, $k_B$ Boltzmann constant, $\mu_\mathrm{B}$ Bohr magneton, $g$ the $g$-factor, and $\chi_0$ the linear slope. Fitting the data yields the results listed in table~\ref{tab1}. The number of quasi-free spins obtained from this analysis through the parameter $M_s$ is again in the $\lesssim1$~\textperthousand-regime if assuming localised moments with $S=1$. The obtained values of $\chi_0$ are rather high with respect to the bare Sommerfeld model: Using $\gamma = 14.5 (13.3)$~mJ/(mol\,K$^2$) for the $Bmab$ ($P2_1/a$) structured material~\cite{PRM2020LNO}, for non-correlated conduction electrons, one obtains $\chi^{Bmab}_{\rm P}\simeq 2.0\times 10^{-4}$~erg/(G$^2$mol) ($\chi^{P2_1/a}_{\rm P}\simeq 1.8\times 10^{-4}$~erg/(G$^2$mol)). Strong anisotropy of $\partial M/\partial B$, at $B=7$~T, already implies that magnetism of conduction electrons is however only one component determining the magnetic response in \lno .

\subsection{Specific heat and \bm{$\partial T_{MM}/\partial B$}}

\begin{figure}[tb]
\centering
\includegraphics[width=\columnwidth,clip]{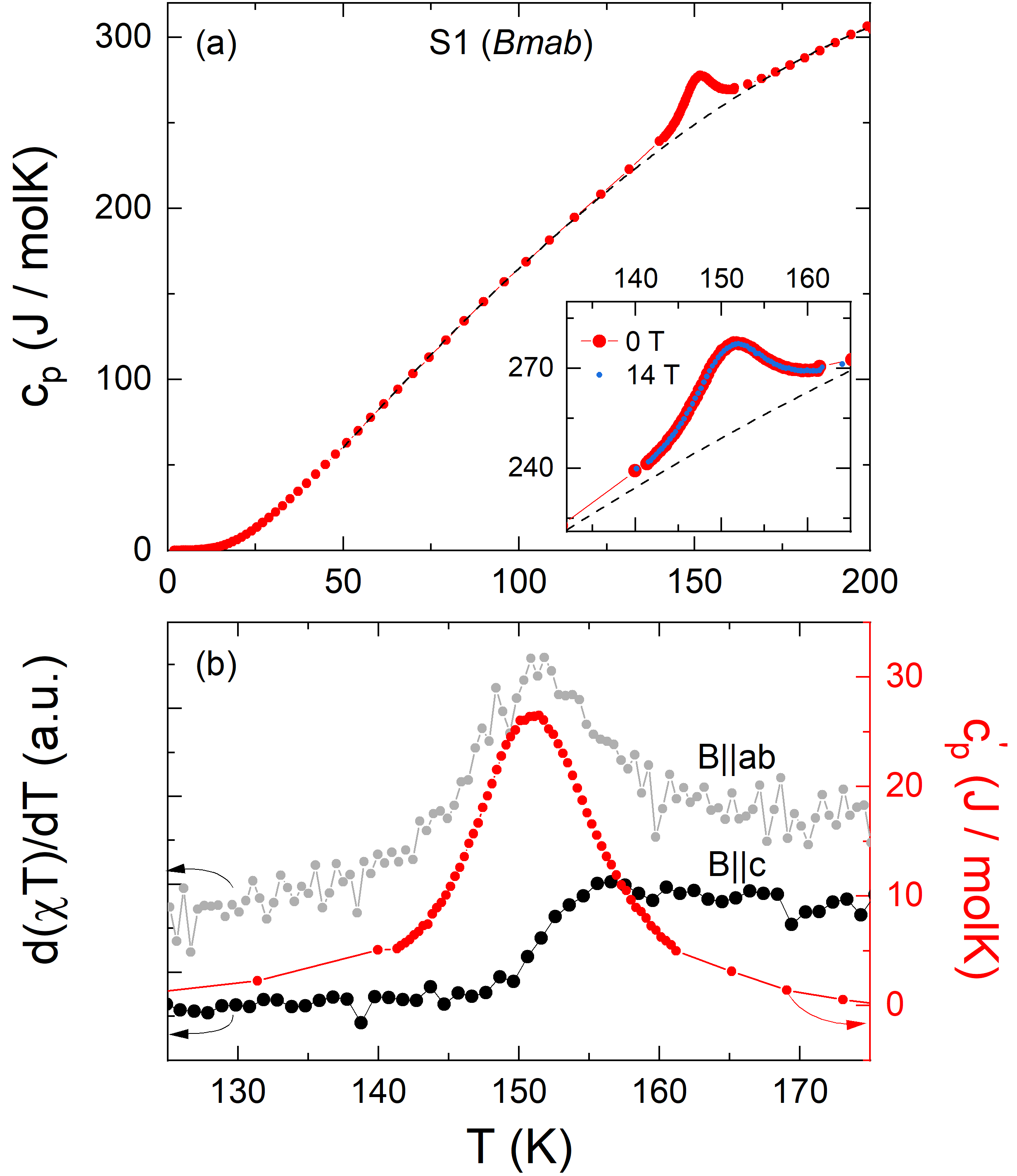}
\caption{(a) Specific heat capacity of the pristine crystal (S1:{\it Bmab}). (b) Fisher's specific heat $\partial (\chi T)/\partial T$ and anomalous contributions to the specific heat $c_{\rm p}'$ obtained by subtracting a polynomial background from the data (dashed line in (a)).}\label{cps1}
\end{figure}

\begin{figure}[tb]
\centering
\includegraphics[width=\columnwidth,clip]{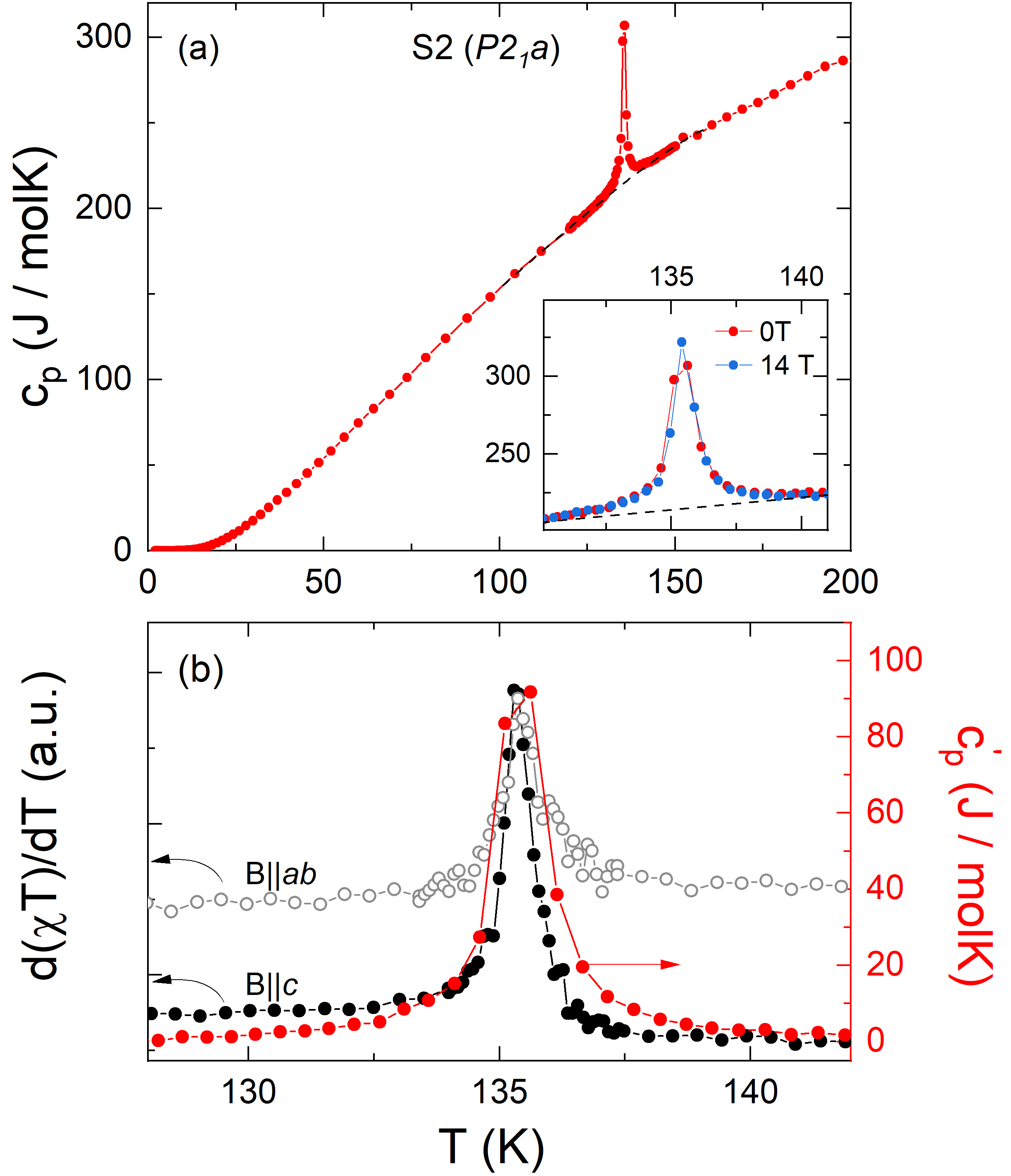}
\caption{(a) Specific heat capacity of the annealed crystal (S2:$P2_1/a$). (b) Fisher's specific heat $\partial (\chi T)/\partial T$ and anomalous contributions to the specific heat $c_{\rm p}'$ obtained by subtracting a polynomial background from the data (dashed line in (a)).}\label{cps2}
\end{figure}

The specific heat of the single crystals under study shown in Fig.~\ref{cps1} and \ref{cps2} confirms significant anomalous entropy changes at \tmmt . While the anomaly of the $P2_1/a$ structured crystal is typical for a first order transition, we conclude from the shape of the anomaly observed for the {\it Bmab} structure the discontinuous character of the MMT, too (see Fig.~\ref{cps1}b). In an attempt to quantify the entropy changes
$\Delta S_{\rm MM}$ associated with the MMT, a polynomial background was fitted to the data well below and above the specific heat anomaly as shown in Fig.~\ref{cps1}a and \ref{cps2}a ~\cite{klingeler2002}. The background mainly reflects the phonon contribution. Due to the large size of the anomaly, using  different temperature ranges for the determination of the background and/or choosing different fit functions does not change the result significantly. Subtracting the obtained background from the data yields the anomaly contribution to the specific heat \Cp ' as shown in Fig.~\ref{cps1}b and \ref{cps2}b. Integrating \Cp '/$T$ yields the entropy changes $\Delta S_{\rm MM}$ listed in table~\ref{tab2}.

Our analysis implies that for both structures, $Bmab$ and $P2_1/a$, the metal-to-metal transition is associated with similar entropy changes. The data also enable us to conclude about the field dependence of \tmmt\ by exploiting the Clausius-Clapeyron equation
\begin{equation}
    \frac{\partial T_{\rm MM}}{\partial B}=-\frac{\Delta M}{\Delta S}. \label{clausius}
\end{equation}
Using the experimentally obtained jumps in $M$ and $S$ yields insignificant effects of magnetic fields on \tmmt\ (see Table~\ref{tab2}) which is experimentally confirmed by our measurements of the specific heat at $B=14$~T (see the insets in Fig.~\ref{cps1}a and \ref{cps2}a).

\begin{figure}[tb]
\centering
\includegraphics[width=0.9\columnwidth,clip]{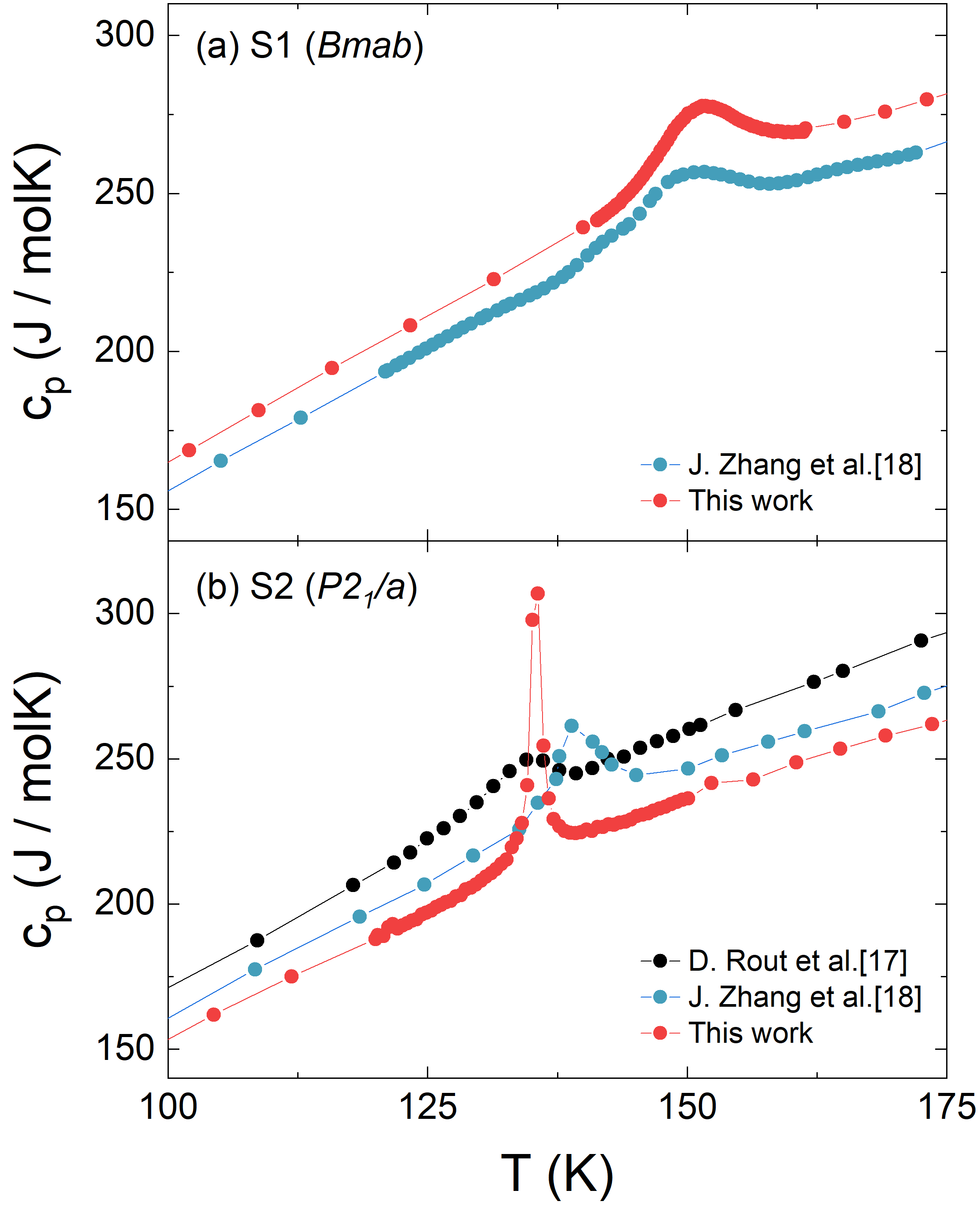}
\caption{Comparison of specific heat capacity of (a) the pristine crystal (S1:$Bmab$) and (b) the annealed crystal (S2:$P2_1/a$) with single crystal~\cite{PRM2020LNO} and polycrstal~\cite{PRB2020RNO-RK} data reported in the literature.}\label{cpcom}

\end{figure}

\begin{table}[h]
\centering
\caption{Changes in entropy ($\Delta S_{\rm MM}$), magnetization ($\Delta M$, and $\Delta (\partial M/\partial T)_{\rm MM}$) at the MMT and field dependencies of \tmmt\ calculated by means of Eq.~\ref{clausius} .}

\begin{tabular}{l|c|c}
  & S1 ($Bmab$) & S2 ($P2_1/a$) \\
\hline
$\Delta S_{\rm MM}$ (\jmk ) &1.4(1)&1.6(1)\\
$\Delta M^{c}_{\rm MM}$ ($10^{-3}$erg/($G^2$\,mol))&-&0.05(1)\\
$\Delta (\partial M/\partial T)^{c}_{\rm MM}$ (erg/$G^2$\,mol\,K)) &-&0.006\\
$\Delta M^{ab}_{\rm MM}$ ($10^{-3}$erg/($G^2$\,mol)) &-0.15(2)&-0.05(1)\\
$|\partial T^{ab}_{\rm MM}/\partial B|$ (mK/T)& $<0.01$&$<0.003$\\

\end{tabular}
\label{tab2}
\end{table}

The obtained entropy changes are by ca. 20~\% smaller than found for single crystal reported in Ref.~\cite{PRM2020LNO} and ca. 40~\% larger than recently determined from a $P2_1/a$-structured polycrystal~\cite{PRB2020RNO-RK}. A comparison of the anomalies is shown in Fig.~\ref{cpcom}. The data also imply slightly different ordering temperatures \tmmt\ which is by 4.5~K lower for $Bmab$ and 2.6~K larger for $P2_1/a$ in Ref.~\cite{PRM2020LNO} than found in our crystals. \tmmt\ of the polycrystal perfectly agrees to our single crystal result~\cite{PRB2020RNO-RK}.  These slight differences may be due to oxygen stoichiometry, which is considered a key parameter of electronic properties~\cite{JAP2000oxygen}. The fact that the specific heat and magnetic susceptibility anomalies at least for the $P2_1/a$ system are much sharper in our crystal as compared to reported anomalies may be regarded as an indication that the crystals have particular excellent crystallinity.

\section{Summary}

\lno\ single crystals were successfully grown and subsequently annealed at 20~bar oxygen pressure. Our specific heat and magnetisation measurements imply $P2_1/a$ and {\it Bmab} structured crystals, respectively, in which the metal-to-metal transitions occur at 152~K and 136~K. Sharp anomalies in the  response functions imply highly crystalline samples and the thermodynamic and magnetic properties are analysed.

\subsection{Acknowledgments}
The authors thank Ilse Glass for technical support. Support by the Deutsche Forschungsgemeinschaft (DFG) under Germany's Excellence Strategy EXC2181/1-390900948 (The Heidelberg STRUCTURES Excellence Cluster) and through project KL1824/13-1 is gratefully acknowledged. N.Y. acknowledges fellowship by the Chinese Scholarship Council (File No. 201906890005).

\subsection{Appendix}

The Supplemental Material contains the full M(B) curve ranging from
-7 T to +7 T.

\bibliography{LNO4310-draft}

\end{document}